\documentclass[12pt]{article}
\usepackage{amssymb}
\usepackage{amsmath}
\usepackage{graphicx}

\begin{document}

\title{NON-MAGNETIC SPINGUIDES AND SPIN TRANSPORT IN SEMICONDUCTORS}
\author{R.N.Gurzhi, A.N.Kalinenko, A.I.Kopeliovich, A.V.Yanovsky\\
B.Verkin Institute for Low Temperature Physics \& Engineering\\
Lenin Ave. 47, Kharkov, 61103 Ukraine\\
 e-mail: gurzhi@ilt.kharkov.ua }
\date{July 27,  2001 г.}
\maketitle
\begin{abstract}
We propose the idea of a "spinguide", i.e. the semiconductor
channel which  is surrounded with walls from the diluted magnetic
semiconductor (DMS) with the giant Zeeman splitting which are
transparent for electrons with the one spin polarization only.
These spinguides may serve as sources of a spin-polarized current
in non-magnetic conductors,  ultrafast switches of a spin
polarization of an electric current and, long distances
transmission facilities of a spin polarization (transmission
distances can exceed a spin-flip length). The selective
transparence of walls leads to new size effects in transport.
\end{abstract}



{\bf 1}. The interest to the spin transport phenomena constantly
grows recently. One of the main problem in this field is creating
a spin polarized current in non-magnetic semiconductors (NS). The
reason is that combining the magnetic data storage with the
electronic readout in one semiconductor device gives a lot of
obvious technological advantages. If, in usual ferromagnetic
metals we have a  spin polarized electric current (in a measure of
a difference  of a density of states for spin-up and spin-down
electrons),  than obtaining  a stationary spin polarization in NS
is fraught with  a number of difficulties. The simple idea is to
pass a current through the interface "ferromagnetic metal -
semiconductor" \cite {Hammar}, but this allows to obtain a
negligible current polarization which is smaller than 0.01 \%
\cite {Molenkamp1}. The reasons of such inefficiency are
following. On the one hand, we have the essential conductivity
mismatch between the ferromagnetic metal and the semiconductor,
and, on the other hand, a spin-flip length is very small in
ferromagnetic metals with a good polarizability.

Recently the new extremely effective method of  spin polarization
of an electric current in NS was realized experimentally \cite
{Molenkamp2}. The sample of diluted magnetic semiconductor $Be _
{x} Mn _ {y} Zn _ {1-x-y} Se$ with a giant splitting of spin
subbands in a magnetic field (the effective g-factor about 100)
has been used as a spin filter in this experiment (for details of
DMS properties  see reviews \cite {Brandt}, \cite {UFN1985}). The
idea of the spin filter is based on the following fact. When the
Fermi level is under a bottom of one of spin subbands in a DMS a
spin-flip scattering is prohibited by virtue of the law of
conservation of energy for low enough temperatures. Besides, the
DMS has a remarkable affinity with the NS used in the experiment
(a semiconductor on the basis of a GaAs), down to coincidens of
lattice parameters. Because a spin-flip process in the NS, being a
relativistic effect, corresponds to a large mean free path (in
some substances up to 100 microns \cite {Molenkamp2}, \cite
{Gantm}), the system of sequentially connected samples of DMS and
NS ideally fits to spin polarization of a current. Accordingly,
almost 100 \% polarization of the current inflowing from DMS into
NS was obtained in the experiment \cite {Molenkamp2}.

From an applied point of view, there is an  important problem of
spins filters application. It is impossible to switch a
polarization very fast because of applying high magnetic field.
Even for not very high magnetic field the delay time is large in
comparison with any mesoscopic time, including the diffusion time.
It is also a technological challenge to create a high
inhomogeneous field with the change of a sign on microscopic
scale. One of the opportunities to obtain fast and compact
switches of a spin polarization, overcoming mentioned problems, is
to use the "non-magnetic spinguides" proposed in the present
paper.

{\bf 2}. {\it Non-magnetic spinguides.} By a "non-magnetic
spinguide" we mean the channel (a  wire or a strip) from a NS with
walls from a DMS (Fig. 1). Unlike the work \cite {Molenkamp2}, in
this case the electric current flows along an interface, instead
of being normal to it. It is supposed, that spinguide walls let
pass electrons with one spin polarization and reflect the another
one. In the absence of a current (in an equilibrium situation)the
NS channel is unpolarized.  However, the current transmitted   by
a spinguide will be polarized, as nonequilibrium electrons with
spins of one of polarization will leave the channel. But for this
purpose it is necessary to earth the DMS walls. Otherwise, if
spin-flip processes are absent, on an exit from the channel there
will be the same unpolarized current, as in the channel entrance.


 Let us consider the diffusion transport regime, when the
diffusion step $\delta l$ is considerably shorter than the channel
width $w$: $\delta l = \min \{ {r_c, l_{i}}\} \ll w$, where $r_c$
is the cyclotron radius, $l_i$ is the electron-impurity mean free
path. (We don't take into account electron-electron collisions,
this is a fortiori valid at temperatures about several Kelvin.)

\begin{figure}[h]
\includegraphics[height=6 cm,width=9 cm]{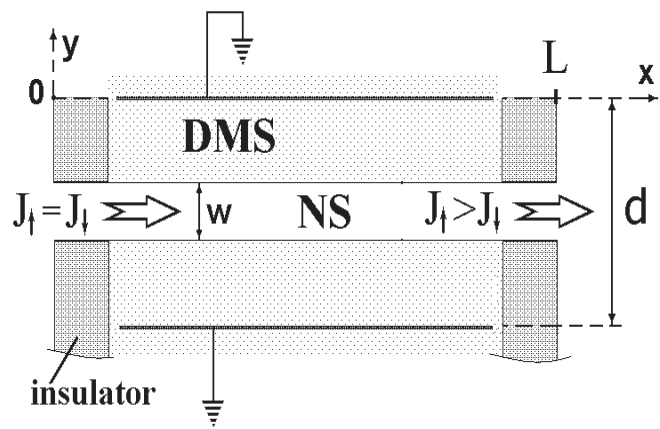}
\caption{The scheme of spinguide. $L$ is the length of the
channel, $w$ -- its width, $d$ -- the distance between the
grounded contacts.}
\end{figure}

Let $\mu_{\uparrow,\downarrow}$ be electrochemical potentials of
carriers with spins up and down, correspondingly. The densities of
electric current are $J_{\uparrow,\downarrow} = - (\sigma/e)
\nabla \mu_{\uparrow,\downarrow}$ (for simplicity the
conductivities $\sigma$ in the NS and the DMS are accepted equal).
For definiteness sake, let us accept that the walls of the
spinguide are transparent for spin-down electrons only, i.e.
$\partial \mu{_\uparrow}/\partial y = 0$ on the boundaries of the
channel (there is supposed to be no dependence on the coordinate
$z$). One should add the earth conditions: $\mu = 0$ on the
grounded contacts. Let non-polarized current $J_{0} =
2J_{\uparrow}(x=0) = 2 J_{\downarrow}(x=0)$ be input. The
diffusion equations $\triangle \mu_{\uparrow,\downarrow} = 0$ can
be solved exactly with taking into account these boundary
conditions. In the case of $L \gg d$, where $d$ is the distance
between the grounded contacts, at the exit of the channel the
current $J_{\downarrow}$ will decrease exponentially with
increasing of the channel length: $J_{\downarrow}\propto\exp\{
-\pi L / d \}$. Taking into account the trivial result
$J_{\uparrow} = J_0/2$, it's easy to see the spin polarization of
a current at the exit from the channel exponentially tends to 1
with increasing of $L$:
\begin{equation}\label{eq4}
  \alpha =
  \frac{J_\uparrow-J_\downarrow}{J_\uparrow+J_\downarrow}\approx 1 - c \cdot \exp \big\{-\frac{\pi
  L}{d}\big\} \ , \ \ c\sim 1 \ .
\end{equation}
It's natural that spin polarization of current $\alpha$ depends
essentially on the geometry of grounded contacts.

It's obvious that the spin polarization in a spinguide is opposite
to that appearing in a spin filter scheme \cite{Molenkamp2} at the
same direction of DMS polarization. Therefore, combining spin
filter and a spinguide with electrostatic gates, it's easy to
switch spin polarization of the current.

Let us note that spin-flip scattering in a NS have a considerably
less influence on a current polarization in a spinguide than in a
spin filter. In the last case a spin polarization of the current
vanishes at the distance $\lambda_S = \sqrt{l_{i} l_S}$ where
$l_S$ is the  mean free path relative to spin-flip processes in a
NS. In a spinguide spin-flip processes manifest themselves in
essentially different way. They affect the polarization weakly (at
$\lambda_S \gg d$) and lead only to reducing of the total current
on the length $\lambda_S$ (for the geometry showed in Fig.1 --
according to exponential law).

\begin{figure}[t]
\includegraphics[height=6 cm,width=9 cm]{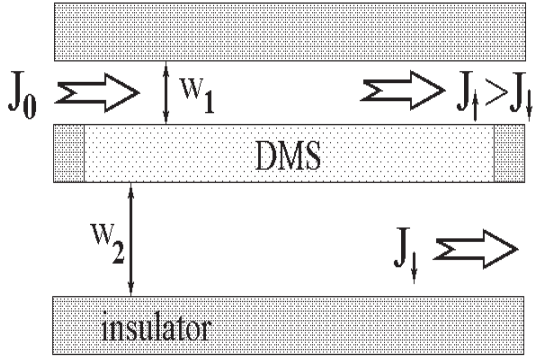}
\caption{Spin drag through a DMS layer. $w_1$, $w_2$ are the
widths of the NS channels.}
\end{figure}

We propose also one more possible design of experiment, namely a
"spin drag" (Fig.2). Non-polarized current is input into channel 1
and this causes entirely polarized current to appear in the
channel 2 (which is separated from channel 1 with DMS layer):
$\alpha = 1$. But the polarization at the exit of channel 1 will
depend on relative width of the channels. In the case of DMS layer
thickness being less than the values $w_1$ and $w_2$ at $L \gg
w_1, w_2$, we have:
\begin{equation}
\alpha =
\frac{1-\gamma}{1+\gamma} \ , \hspace{10 mm} \gamma =
\frac{w_1}{w_1 + w_2} \ .
\end{equation}
The current polarizations in channels 1 and 2 are opposite, but
the integral current of two channels, of course, is non-polarized.
It's interesting that for a wide channel 2, $w_1 \ll w_2 \ll L$,
polarized currents will be equally divided between the channels,
i.e. in the channel 1 we'll have entirely polarized current
$J_\uparrow = J_0/2$, with equal value and opposite polarization
to that in channel 2. It is accepted at the development of formula
2 that the same potential is applied at the exit of channel 2 as
at the exit of channel 1. (The latter in our model is determined
by the value of current $J_0$). Changing the potential at the exit
of channel 2,one can control the current polarization in the
channels.


{\bf 3}. As one can see from the stated above, spinguides let us
achieve a high level of a spin polarization of a current in
non-magnetic conductors. Though the use of spin filters also gives
a high level of polarization, spinguides can be useful in
connection with the opportunity of easy current spin polarization
control and long-distance carrying of polarization, in spite of
spin-flip scattering. Besides, at the conditions of selective
transparency of the boundaries in such systems some interesting
features of galvanomagnetic phenomena can appear. In particular,
let us note that delocalized states in a magnetic field on the
DMS-NS boundaries (corresponding to classic skipping orbits) will
be entirely spin-polarized.



\begin{thebibliography}{0}
\bibitem{Hammar} P. R. Hammar, B. R. Bennet, M.J. Yang and M.Johnson, Phys.Rev.Lett. {\bf 83}, 203 (1999).
\bibitem{Molenkamp1} G. Shmidt, D. Ferrand, L. A. T. Filip, L. W. Molenkamp and B. J. van Wees, Phys.Rev.
B {\bf 62}, R4790 (2000).
\bibitem{Molenkamp2} R. Fiederling, M. Keim, G. Reuscher, W. Ossau, G. Schmidt, A. Waag, L. W. Molenkamp,
Nature {\bf 402}, 787 (1999).
\bibitem{Brandt} N. B. Brandt and V. V. Moschchalkov, Adv.Phys. {\bf 33}, 193 (1984).
\bibitem{UFN1985} I. I. Liapilin, I. M. Cidilkovsky, UFN {\bf 146}, 35 (1985).
\bibitem{Gantm} V. F. Gantmaher, I. B. Levinson,
Current Scattering in Metals and Semiconductors, Nauka, Moscow, 352 p. (1984).
\end{thebibliography}
\end{document}